\documentclass[twocolumn,aps]{revtex4}
\usepackage{graphicx}
\begin{document}
\title{Ionization rates in a Bose-Einstein condensate
of metastable Helium}
\author{O. Sirjean, S. Seidelin, J. Viana Gomes, D. Boiron, C. I. Westbrook and A. Aspect}
\affiliation{\mbox{Laboratoire Charles Fabry de l'Institut
d'Optique, UMR 8501 du CNRS, F-91403 Orsay Cedex, France}}

\author{G. V. Shlyapnikov}
\affiliation{\mbox{FOM Institute for Atomic and Molecular Physics,
Kruislaan 407, 1098 SJ Amsterdam, The Netherlands}\\
Russian Research Center Kurchatov Institute, Kurchatov Square,
123182 Moscow, Russia}

\begin{abstract}
We have studied ionizing collisions in a BEC of He*. Measurements
of the ion production rate combined with measurements of the
density and number of atoms for the same sample allow us to
estimate both the 2 and 3-body contributions to this rate. A
comparison with the decay of the number of condensed atoms in our
magnetic trap, in the presence of an rf-shield, indicates that
ionizing collisions are largely or wholly responsible for the
loss. Quantum depletion makes a substantial correction to the
3-body rate constant.

\end{abstract}
\maketitle

The observation of Bose-Einstein condensation of metastable helium
(He in the $2^{3}S_{1}$ state, denoted He*) \cite{BecHe,BecENS}
constituted a pleasant surprise for experimentalists although the
possibility had been predicted theoretically
\cite{Shlyapnikov:94a}. Success hinged, among other things, on a
strong suppression of Penning ionization in the spin-polarized,
magnetically trapped gas. Too high a rate of ionization would have
prevented the accumulation of a density of atoms sufficiently high
to achieve evaporative cooling. The ionization rate is not
completely suppressed however, and when the atomic density gets
high enough, a cold, magnetically trapped sample of He* does
produce a detectable flux of ions. As shown in \cite{BecHe}, this
signal can even be used as a signature of BEC. The observation of
ions from the condensate opens the possibility of monitoring in
real time the growth kinetics of a condensate \cite{NaissanceBEC}.
This is an exciting prospect, but in order to quantitatively
interpret the ion rate, it is necessary to know the relative
contributions of 2 and 3-body collisions.

In this paper we use the unique features of metastable atoms to
detect, in a single realization of a BEC, the ionization rate, the
density and the number of atoms. This allows us to extract 2 and
3-body rate constants without relying on fits to non-exponential
decay of the number of atoms, which require good experimental
reproducibility \cite{Roberts,3corpsDalibar,Burt} and are
notoriously difficult to interpret quantitatively \cite{Roberts}.
After estimating the ionization rate constants, a comparison with
the observed decay of the number of atoms reveals no evidence for
collisional avalanche processes. Thus, by contrast with $^{87}$Rb
\cite{Rempe}, He$^{*}$ seems to be a good candidate for studying
``hydrodynamic" regime as well as the effects of quantum
depletion. Indeed in our analysis of the 3-body ionization
process, quantum depletion makes a substantial correction
\cite{DepletionGora}.

Much theoretical \cite{Shlyapnikov:94a,venturi:99} and
experimental \cite{Hill:72,Vassen,BecHe,BecENS} work has already
been devoted to estimating inelastic decay rates in He*. The
dominant 2-body decay mechanisms, which we will refer to
collectively as Penning ionization,
\begin{equation}\label{2corps}
He^{*} + He^{*} \rightarrow \left \{
\begin{array}{l}
He^{+} + He(1S) + e^{-} \\
He_{2}^{+} + e^{-}
\end{array}
\right.
\end{equation}
are known to be suppressed by at least 3 orders of magnitude in a
spin-polarized sample, but the total rate constant has not yet
been measured. The theoretical estimate of the rate at 1 $\mu$K is
$\sim 2 \times 10^{-14}\textrm{ cm}^{3}\textrm{ s}^{-1}$
\cite{Shlyapnikov:94a,venturi:99}. The 3-body reaction,
\begin{equation}\label{3corps}
\begin{array}{rccl}
He^{*} + He^{*} + He^{*}& \rightarrow & He_{2}^{*}& + He^{*} (\sim
1mK) \\
 & & \hookrightarrow & He^{+} + He(1S) + e^{-}
\end{array}
\end{equation}
proceeds via 3-body recombination followed by autoionization of
the excited molecule. The rate has been estimated theoretically
\cite{Shlyapnikov:96b} to have a value of order
$10^{-26}$~cm$^{6}$~s$^{-1}$. Both reactions yield one positive
ion which can easily be detected in our apparatus.

We define collision rate constants according to the density loss
in a thermal cloud \cite{notedefcoef}: $ \frac{dn}{dt} = -
\frac{n}{\tau} - \beta \, n^{2} - L \, n^{3} $ with $n$ the local
density, $\tau$ the (vacuum limited) lifetime of the sample, and
$\beta$ and $L$ the 2-body and 3-body ionizing rate constants
defined for a thermal cloud. We have assumed here that there are
no other loss processes. One can calculate an expected ionization
rate per trapped atom ($\Gamma$):
\begin{equation}\label{formulegamma}
 \Gamma=\frac{Ion~rate}{ N_{0}} = \frac{1}{\tau'} +
\frac{2}{7} \; \kappa_{2} \; \beta \; n_{0} +
\frac{8}{63} \; \kappa_{3} \; L \;  n_{0}^{2},
\end{equation}
for a pure BEC in the Thomas-Fermi regime with a number of atoms
$N_{0}$, and a peak density $n_{0}$. The numerical factors come
from the integration over the parabolic spatial profile and the
fact that although 2 or 3 atoms are lost in each type of
collision, only 1 ion is produced. The effective lifetime $\tau'
\geq \tau$ is due to ionizing collisions with the background gas.
The factors $\kappa_{i}$ take into account the fact that the 2 and
3-particle local correlation functions are smaller than those of a
thermal cloud. For a dilute BEC $\kappa_{2}=1/2!$ and
$\kappa_{3}=1/3!$ \cite{DepletionGora,Burt}. Because the He*
scattering length is so large, quantum depletion lead to
significant corrections \cite{DepletionGora} to the $\kappa$'s as
we discuss below.

Much of our setup has been described previously
\cite{BecHe,Therma,Nowak:00}. Briefly, we trap up to $2\times
10^8$ atoms at 1~mK in a Ioffe-Pritchard trap with a lifetime
($\tau$) of 90~s. We use a "cloverleaf" configuration
\cite{Mewes:96} with a bias field $B_0 =150$~mG. The axial and
radial oscillation frequencies in the harmonic trapping potential
are $\nu_{\parallel}=47\pm3$~Hz and $\nu_{\bot}=1800\pm50$~Hz
respectively
($\overline{\omega}/2\pi=(\nu_{\parallel}\nu_{\perp}^{2})^{1/3}=534$~Hz).
A crucial feature of our set up is the detection scheme, based on
a 2 stage, single anode microchannel plate detector (MCP) placed
below the trapping region. Two grids above the MCP allow us either
to repel positive ions and detect only the He* atoms, or to
attract and detect positive ions produced in the trapped cloud.

To detect the ion flux, the MCP is used in counting mode: the
anode pulses from each ion are amplified, discriminated with a
600~ns deadtime and processed by a counter which records the time
delay between successive events. Typical count rates are between
$10^2$ and $10^4$~s$^{-1}$. We have checked that the correlation
function of the count rate is flat, indicating that there is no
double counting nor any significant time correlation in the ion
production. The dark count rate is of order 1~s$^{-1}$. By
changing the sign of the grid voltage, we have checked that while
counting ions, the neutral He* detection rate is negligible
compared to the ion rate (less than 5~\%) even when the radio
frequency (rf) shield is on. We estimate the ion detection
efficiency by assuming that only ions which hit the open channels
of the MCP (60\% of the total area) are detected (with a 100\%
quantum efficiency). We then multiply by the transmission of the
two grids $(0.84)^{2}$. Based on Refs. \cite{Deconihout,MCP}, we
assume this (0.42) is an upper limit on our detection efficiency.

To find the values of $N_0$ and $n_0$ corresponding to the
measured ion rate, we use the MCP to observe  the time-of-flight
signal (TOF) of the He* atoms released from the rapidly switched
off trap. The instantaneous count rate can be as high as
$10^6$~s$^{-1}$, and the MCP saturates when used in counting mode.
To avoid this problem, we lower the MCP gain, and record the TOF
signal in analog mode with a time constant of 400 microseconds.
Several tests were performed to verify the linearity of the
detector.

\begin{figure}
\begin{center}
\includegraphics[height=5cm]{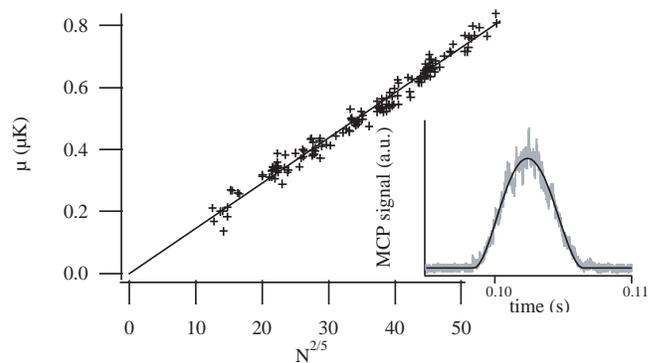}
\end{center}
\caption{Chemical potential versus number of detected atoms to the
power $\frac{2}{5}$ and its linear fit. Data are for quasi-pure
BEC. Inset shows a typical TOF signal and its inverted parabola
squared fit. }\label{figMuNo25}
\end{figure}

In a typical run, forced evaporative cooling takes place for 40~s,
down to an rf-knife frequency of 500~KHz, about 50~kHz above the
minimum of the trapping potential. Near the end of the ramp, the
ion rate increases sharply, signaling the appearance of a BEC
(Fig. 4 in \cite{BecHe}). After reaching the final value, the
rf-knife is held on at that frequency (rf-shield). The above
sequence results in a quasi pure BEC for delay times up to 15~s
(see Fig. \ref{decayNo}). By quasi-pure we mean that we see no
evidence of any thermal wings in signals such as shown in the
inset of Fig. \ref{figMuNo25}. From tests of our fitting
procedure, we estimate that the smallest thermal fraction we can
distinguish is about 20~\%, corresponding to a temperature on the
order of the chemical potential. Runs in which thermal wings were
visible were discarded.

To acquire the TOF signals corresponding to a given ion rate, we
turn off the rf-shield, wait 50~ms, and then turn off the magnetic
trap, and switch the MCP to analog mode. To be sure that the rf
has no influence on the ion rate, we use only the number of ions
observed during the 50~ms delay to get the rate. We fit the TOF
signals to an inverted parabola squared as expected for a pure BEC
in the Thomas-Fermi regime, and for a TOF width ($\sim 5$~ms)
narrow compared to the mean arrival time ($100$~ms) \cite{BecHe}.
Under these assumptions, the chemical potential $\mu$ depends only
on the TOF width, the atomic mass and the acceleration of gravity
\cite{Castin}, and thus can be measured quite accurately. Figure
\ref{figMuNo25} shows that $\mu$ varies as $N_0^{2/5}$ as
expected, over almost 2 decades in atom number. Residuals from the
linear fit do not show any systematic variation which is a good
indication of the detection linearity. A fit on a log-log plot
gives a slope of $0.39$.

To determine the collision rate constants $\beta$ and $L$, we need
an absolute calibration of the number of atoms and the density. As
discussed in Ref. \cite{BecHe}, all the atoms are not detected,
and the direct calibration has a 50~\% uncertainty which is
responsible for the large uncertainty in the scattering length
$a$. In fact the measurement of the chemical potential gives an
accurate value for the product $n_{0} \, a =\mu m /4\pi\hbar^{2}$,
and with the value of $\overline{\omega}$ gives the product $N_{0}
\, a =(1/15) \, (\hbar / m \overline{\omega})^{1/2}\, (2\mu /
\hbar\overline{\omega})^{5/2} $ as well. Therefore, in the hopes
that the He* scattering length will be measured more accurately in
the future, we shall express $N_0$ and $n_0$ in terms of $a$. In
this paper, unless stated otherwise, we suppose that $a=20 \,
\textrm{nm}$, and in our conclusions we shall discuss how our
results depend on $a$.

\begin{figure}
\begin{center}
\includegraphics[height=5cm]{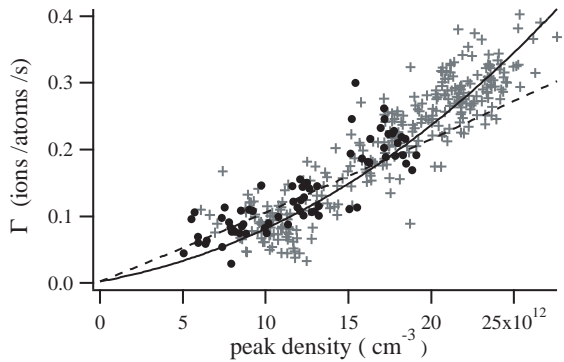}
\end{center}
\caption{Ion rate per trapped atom versus peak density for 350
different quasi-pure BEC's. Atom number and density are deduced
from $\mu$ , $\overline{\omega}$ and $a$ (here 20~nm). Data were
taken for 2 different bias fields corresponding to $\nu_{\bot} =
1800$~Hz (crosses) and $\nu_{\bot} = 1200$~Hz (circles). The
dashed line corresponds to the best fit involving only 2-body
collisions. The solid line is a fit to 2 and 3-body processes.}
\label{figresults}
\end{figure}

\begin{figure}
\begin{center}
\includegraphics[height=5cm]{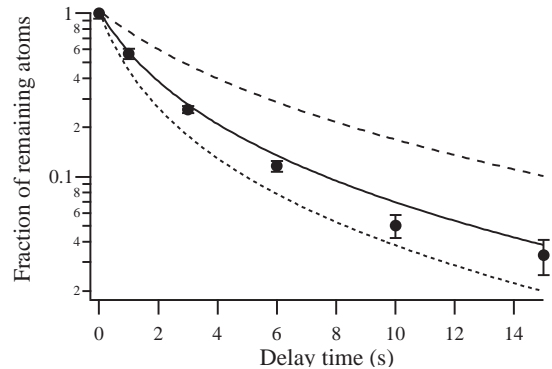}
\end{center}
\caption{ Fraction of remaining atoms measured by TOF as a
function of time. The rf-shield is on and the cloud remains a
quasi pure condensate during the decay. The lines correspond to
the predicted atom decay according to Eq. 3 with the fitted value
of the 2- and 3-body rate constants for $a=10$~nm (dashed line),
$a=20$~nm (solid line) and $a=30$~nm (dotted line). The case of
$a=10$~nm is not necessarily excluded because other, non-ionizing
losses could be present.} \label{decayNo}
\end{figure}

Figure \ref{figresults} shows the ion rate per atom $\Gamma$
\textit{versus} the peak density.  The densest sample corresponds
to $N_{0}=2 \times 10^{5}$ atoms and $n_{0}=2.5 \times
10^{13}$~cm$^{-3}$. The corresponding Thomas-Fermi radii are
$r_\bot \simeq 5~\mu$m and $r_\parallel \simeq 200~\mu$m. The
vertical intercept in Fig. \ref{figresults} corresponds to
ionizing collisions with the background gas ($1/\tau'$). We have
independently estimated this rate using trapped thermal clouds at
1~mK and 5~$\mu$K, and found $1/\tau' \lesssim 5 \times
10^{-3}$~s$^{-1}$. This value is negligible at the scale of the
figure.

The curvature seen in Fig. \ref{figresults} shows that 3-body
ionizing collisions are significant. Before fitting the data to
get $\beta$ and $L$, we must take into account several effects.
First, for 3-body collisions, quantum depletion is important. For
$T=0$, reference \cite{DepletionGora} gives a multiplicative
correction \cite{CorrectionGora} to the factor $\kappa_{3}$ of
$((1+\epsilon)=(1+ A \frac{72}{\sqrt{{\pi}}} \times \sqrt{n_{0} \,
a^{3}}))$, where $A\simeq 0.84$ and comes from an integration over
the spatial profile using a local density approximation. At our
highest density $\epsilon \simeq 0.5$. Two-body collisions are
subject to an analogous correction but approximately 3 times
smaller. The fits in Fig. \ref{figresults} include the density
dependence of $\kappa_{2,3}$, associated with quantum depletion.
The $n_0^{3/2}$ dependence introduced for 2-body collisions is far
too small to explain the curvature in the data. The density
dependence of $\kappa_{2,3}$ does not significantly improve the
quality of the fit, but it significantly affects the value of the
fitted value of $L$ (reduction of 40\%).

In addition, the fact that the sample probably contains a small
thermal component means that collisions between the condensed and
the thermal parts must be taken into account
\cite{DepletionGora,3corpsDalibar}. Assuming a 10~\% thermal
population ($\frac{\mu}{k_{b}T}\simeq1.1$), we find $\kappa_{3} =
\frac{1}{6}(1+\epsilon+ \epsilon')$, with an additional correction
$\epsilon'\simeq 0.35$ for the densest sample
\cite{QuantumDepletionT}.

Taking into account all these corrections, the fitted values of
the collision rate constants \cite{notedefcoef} are: $ \beta_{20}
= 2.9 (\pm2.0) \times 10^{-14}$~cm$^{3}$ sec$^{-1}$ and $L_{20} =
8.5 (\pm5.3) \times 10^{-27}$~cm$^{6}$ sec$^{-1}$, where the
subscripts refer to the assumed value of $a$. These values are in
good agreement with the theoretical estimates. The error bars are
estimated as follows. We fix either $\beta$ or $L$ and use the
other as a fit parameter. We repeat this procedure for different
values of the fixed parameter and take the range over which we can
get a converging and physically reasonable fit (i.e. no negative
rate constants) as the uncertainty in the fixed parameter. These
error bars are highly correlated since if $\beta$ is increased,
$L$ must be decreased and vice-versa. The error bars do not
include the uncertainty in the absolute ion detection efficiency
(see below).

Until now we have assumed $a=20$~nm, but current experiments give
a range from 8~nm to 30~nm \cite{BecHe,BecENS}. Using Eq. 3 and
our parameterization of $n_{0}$ and $N_{0}$ in terms of $a$, one
can see that, in the absence of quantum depletion, the values of
$\beta$ and $L$ extracted from our analysis would be proportional
to $a^{2}$ and $a^{3}$ respectively. Taking quantum depletion into
account, no simple analytical dependence exists, but one can
numerically evaluate $\beta$ and $L$ \textit{vs.} $a$ and fit the
results to expansions with leading terms in $a^{2}$ and $a^{3}$
respectively. The effect of quantum depletion is negligible for
$\beta$ ($\beta_{a} \approx \beta_{20} (\frac{a}{20})^{2}$). For
$L$, we find $ L_a\approx L_{20}(\frac{a}{20})^{3}[1 -
0.23\frac{a-20}{20}]$ with $a$ in nm.

To test the consistency of our measurements, we have studied the
decay of the number of atoms in the BEC (Fig. \ref{decayNo}). To
acquire these data, we held the BEC in the trap in the presence of
the rf-shield for varying times. This study involves multiple
realizations of a BEC, which typically exhibit large fluctuations
in the initial atom number. We have been able to reduce this noise
by using the ion signal to select only data corresponding to the
same ion rate 500~ms after the end of the ramp. This time
corresponds to $t=0$ in the figure. We have also plotted the
predicted decay curve (solid line) corresponding to ionization
only. This curve results from a numerical integration of the atom
loss due to ionization processes, calculated from the fitted
values $\beta_{20}$ and $L_{20}$. The fact that the error bars on
$\beta$ and $L$ are correlated leads to a small uncertainty on the
solid curve that happens to be of the same order of magnitude as
the typical error bars on the data. The observed decay agrees
fairly well with the solid curve, and ionization apparently
accounts for most of the loss. If the ion detection efficiency
were actually lower than we assume, the predicted decay due to
ionization would be faster than the observed decay, an unphysical
situation. From this, we conclude that our estimate of the
detection efficiency is accurate and does not lead to an
additional uncertainty in $\beta$ and $L$.

We have also plotted the curves obtained from the same analysis
but with scattering lengths of 10 and 30~nm. The curve
corresponding $a=30$~nm lies below the data points. Based on our
analysis, this means that $a=30$~nm is excluded. A scattering
length of $a=25$~nm is the largest one consistent with our data.
In contrast, the decay predicted for an analysis with
$a=10\textrm{~nm}$ is slower than the observed decay of the number
of condensed atoms. This would mean that there are additional
non-ionizing losses (contributing up to half of the total loss),
and/or that we have overestimated the ion detection efficiency by
a factor as large as 2. In the latter case, the rate constants
$\beta$ and $L$ should be multiplied by the same factor. This
results for $a=10~$nm in a supplementary systematic uncertainty on
$\beta$ and $L$ of a factor as large as 2.

Even though the peak densities of our BEC are small compared to
those in alkalis, the elastic collision rate is high because of
the large scattering length, and one must consider the possibility
of collisional avalanches. For $a=20$~nm our densest cloud has a
mean free path of $(8\pi a^2\overline{n})^{-1}\approx 7\mu$m and
using the definition of \cite{Rempe} the collisional opacity is
$\approx 0.8$. With Rb atoms this would result in avalanche
processes increasing by a large amount the atom loss \cite{Rempe}.
In our case we have to consider secondary collisions leading to
both ion production and atom loss. However, no product of an
ionizing collision can produce many secondary ionizing collisions,
since the corresponding mean free paths are at least two orders of
magnitude larger than $r_\parallel$. Hence secondary ionization is
unimportant and Eq. \ref{formulegamma} correctly describes the
ionizing processes. This conclusion is supported by our
observation that no correlation exists in the time distribution of
detected ions.

The good agreement between the data and the curve in Fig. 3
indicates that losses due to non-ionizing collisional avalanches
are not taking place either. This is in agreement with data on
elastic collisions with He$^{+}$, $He_{2}^{+}$ and He$(1S)$, which
have small cross sections \cite{Beijerinck}. Collisions with hot
He* atoms from the reaction of Eq. \ref{3corps} are more likely to
play a role, but due to the higher velocity, the elastic cross
section for these atoms is smaller. This is in contrast to what
happens with Rb atoms \cite{Rempe} where the total cross section
is enhanced due to d-wave scattering resonance.

The theoretical analysis shows that quantum depletion strongly
affects the measured 3-body rate constant. One way to
experimentally demonstrate this effect would be to do similar
measurements with thermal clouds, and compare them with the
results reported here. Absolute calibration of ion and atom
detection efficiency should play no role in this comparison, if
one could prove that they are the same for both situations.

We thank F. Gerbier for stimulating discussions. Supported by the
European Union under grants IST-1999-11055 and HPRN-CT-2000-00125,
and by the DGA grant 00.34.025.


\begin{thebibliography}{2}

\bibitem{BecHe}
A. Robert {\it{et al.}}, Science {\bf 292}, 461 (2001).

\bibitem{BecENS}
F. Pereira Dos Santos {\it{et al.}}, Phys. Rev. Lett. {\bf 86},
3459 (2001).

\bibitem{Shlyapnikov:94a}
G. V. Shlyapnikov {\it{et al.}}, Phys.~Rev.~Lett. {\bf 73}, 3247
(1994); P. O. Fedichev {\it{et al.}}, Phys.~Rev.~A {\bf 53}, 1447
(1996).

\bibitem{NaissanceBEC}
H. J. Miesner {\it et al.}, Science, {\bf 270}, 1005 (1998); M.
K\"{o}hl {\it et al.}, Phys. Rev. Lett. {\bf 88}, 080402 (2002).


\bibitem{Roberts}
J. L. Roberts {\it et al.}, Phys. Rev. Lett. {\bf 85}, 728 (2000).

\bibitem{3corpsDalibar}
J. S\"{o}ding {\it et al.}, Appl. Phys. B {\bf 69}, 257 (1999).

\bibitem{Burt} E. A. Burt {\it et al.}, Phys. Rev. Lett. {\bf 79},
337 (1997).

\bibitem{Rempe}
J. Schuster {\it et al.}, Phys. Rev. Lett. {\bf 87}, 170404
(2001).

\bibitem{DepletionGora}
Y. Kagan, B. V. Svistunov, and G. V. Shlyapnikov, JETP Lett. {\bf
42}, 209 (1985).

\bibitem{venturi:99}
V. Venturi {\it{et al.}}, Phys. Rev. A {\bf 60}, 4635 (1999); V.
Venturi and I. B. Whittingham, Phys. Rev. A {\bf 61}, 060703(R)
(2000).

\bibitem{Hill:72}
J. C. Hill {\it{et al.}}, Phys.~Rev.~A {\bf 5}, 189 (1972).

\bibitem{Vassen}
N. Herschbach {\it{et al.}}, Phys.~Rev.~A {\bf 61}, 50702 (2000).

\bibitem{Shlyapnikov:96b}
P. O. Fedichev, M. W. Reynolds, and G. V. Shlyapnikov, Phys. Rev.
Lett. {\bf 77}, 2921 (1996); P. F. Bedaque, E. Braaten, and H. W.
Hammer, Phys.~Rev.~Lett. {\bf 85}, 908 (2000).

\bibitem{notedefcoef}
Collision rate constants are sometimes defined directly for a BEC
($\beta'=\beta/2$ and $L'=L/6$).

\bibitem{Therma}
A. Browaeys {\it{et al.}}, Phys. Rev. A {\bf 64}, 034703 (2001).

\bibitem{Nowak:00}
S. Nowak {\it et al.}, Appl. Phys. B {\bf 70}, 455 (2000).

\bibitem{Mewes:96}
M. O. Mewes {\it et al.}, Phys.~Rev.~Lett. {\bf 77}, 416 (1996).

\bibitem{Deconihout}
B. Deconihout {\it et al.}, Appl. Surf. Sci. {\bf 94/95}, 422
(1996).

\bibitem{MCP}
R. S. Gao {\it et al.}, Rev. Sci. Instrum. {\bf 55}, 1756 (1984).

\bibitem{Castin}
Y. Castin and R. Dum, Phys. Rev. Lett. {\bf 77}, 5315 (1996); Y.
Kagan, E.L. Surkov, and G. V. Shlyapnikov, Phys. Rev. A {\bf 54},
1753 (1996).

\bibitem{CorrectionGora}
The numerical factor 64 of \cite{DepletionGora} is replaced by a
factor 72 because here we express the decay rate in terms of the
condensate density $n_{0}$, not in terms of the total density.

\bibitem{QuantumDepletionT}
For $T \sim \mu$, we have $\epsilon' \approx
<3/\pi^{2}\int_{0}^{\infty}
(\frac{3E_{k}}{2\varepsilon_{k}(r)}+\frac{\tilde{\mu}(r)}{2\varepsilon_{k}(r)})\frac{1}{e^{\varepsilon_{k}(r)/k_{b}T}-1}
k^{2} dk / n(r) > $ with $n(r)$ the local density of the BEC,
$E_{k}=\frac{\hbar^{2}k^{2}}{2m}$,
$\varepsilon_{k}(r)=\sqrt{E_{k}^{2}+2 \tilde{\mu}(r) E_{k}}$, and
$\tilde{\mu}(r)$ the local chemical potential. The brackets
indicate spatial averaging. This can be written : $48 \,
\sqrt{\frac{2}{\pi}} \times \sqrt{n_{0}a^{3}} \times
A'(\frac{\mu}{k_{b}T})$. The numerical factor $A'$ comes from the
integration over the spatial profile and takes implicitly into
account the overlap of the BEC with the thermal cloud. We find
$A'(\frac{\mu}{k_{b}T}=1.1) \simeq 0.65$.



\bibitem{Beijerinck}
H. C. W. Beijerinck {\it et al.}, Phys. Rev. A {\bf 61}, 23607
(2000).

\end{thebibliography}
\end{document}